# Force Feedback Effects on Single Molecule Hopping and Pulling Experiments

M. Rico-Pasto[1], I. Pastor[1,2] and F. Ritort[1,2]

[1]Departament de Fisica de la Materia Condensada, Universitat de Barcelona, C/ Marti i Franques 1, 08028, Barcelona, Spain

[2]CIBER_BNN, Instituto de Salud Carlos III, 28029, Madrid, Spain

**ABSTRACT**

Single-molecule experiments with optical tweezers have become an important tool to study the properties and mechanisms of biological systems, such as cells and nucleic acids. In particular, force unzipping experiments have been used to extract the thermodynamics and kinetics of folding and unfolding reactions. In hopping experiments, a molecule executes transitions between the unfolded and folded states at a preset value of the force (constant force mode -CFM- under force feedback) or trap position (passive mode -PM- without feedback) and the force-dependent kinetic rates extracted from the lifetime of each state (CFM) and the rupture force distributions (PM) using the Bell-Evans model. However, hopping experiments in the CFM are known to overestimate molecular distances and folding free energies for fast transitions compared to the response time of the feedback. In contrast, kinetic rate measurements from pulling experiments have been mostly done in the PM while the CFM is seldom implemented in pulling protocols. Here, we carry out hopping and pulling experiments in a short DNA hairpin in the PM and CFM at three different temperatures (6ºC, 25ºC and 45ºC) exhibiting largely varying kinetic rates. As expected, we find that equilibrium hopping experiments in the CFM and PM perform well at 6ºC (where kinetics is slow) whereas the CFM overestimates molecular parameters at 45ºC (where kinetics is fast). In contrast, nonequilibrium pulling experiments perform well in both modes at all temperatures. This demonstrates that the same kind of feedback algorithm in the CFM leads to more reliable determination of the folding reaction parameters in irreversible pulling experiments.

## I. INTRODUCTION

The invention of single molecule manipulation techniques over the past decades has provided new insights into the details of complex molecular reactions in cells [1-5] that complement traditional bulk methods. Techniques such as optical tweezers [6,7] allow scientists to mechanically unzip and stretch single biomolecules like DNA [8-10], RNA [11,12] and proteins [13-15] in a controlled manner.

In DNA unzipping experiments a tensile force is applied to the 3' and 5' of a DNA hairpin until the base pairs that stabilize the double helix are disrupted and the native hairpin unfolds and is converted into single-stranded DNA (ssDNA) [10,16,17]. The unzipping experiment is the equivalent of a temperature-induced melting process, the main difference being the final state of the DNA hairpin: a stretched polymer in the former versus a random coil in the latter. The reverse of the unzipping process is molecular folding or zipping: upon releasing the force, the stretched ssDNA folds back into the native hairpin in a process determined by the nucleation of the loop and the formation of the stem. The thermodynamics and kinetics of the unzipping-zipping reaction gives valuable information about the free energy of folding of the DNA hairpin and the underlying molecular free energy landscape otherwise difficult to obtain in bulk assays. The unfolding and folding reaction can be studied in two types of force spectroscopy protocols using optical tweezers: hopping and pulling. Moreover, experiments can be carried out in two control modes: constant-force mode (CFM) and passive mode (PM) depending on whether the force (CFM) or the trap position (PM) is the control parameter (i.e. the externally controlled variable that can be held fixed and is not subject to thermal fluctuations).

In hopping experiments, the control parameter (trap position or force) is kept constant while the molecule executes transitions between the folded and unfolded states [18,18-21]. In contrast, in pulling experiments the control parameter is repeatedly and continuously changed between two limit values while the force-distance curve is recorded along the stretching-releasing cycles [22,23]. Single molecule experiments monitor transitions in real time, allowing us to measure transiently fast and rare events that would otherwise be masked in bulk assays where only averages over a large number of molecules are measured.

The force-dependence of the measured unfolding and folding kinetic rates are typically studied using the Bell-Evans (BE) model [24,25] where the unfolding-folding process is a thermally activated diffusive process that passes through a transition state characterized by a kinetic barrier. Kinetic rates exhibit an Arrhenius-like exponential dependence where the kinetic barrier in the exponent changes linearly with force. Fitting the BE model to the experimentally measured kinetic rates allows us to estimate the distance from each state (folded and unfolded) to the transition state [20,26]. The BE model is based on the force ensemble, i.e. force is the control parameter in the experiment. In an optical tweezers setup, however, the position of the center of the optical trap rather than force is the natural control parameter. Because kinetics is directly dependent on force scientists have made efforts to implement algorithms of force feedback to control directly the force in hopping experiments [18,19]. However, force feedback algorithms present artifacts due to their finite bandwidth that lead to limited finite response time of the device and additional low frequency colored noise acting upon the system. The first artifact leads to fast transitions being missed during the algorithm feedback cycle and an overestimation of the molecular extension as obtained from kinetic rates measurements [19].

In this work we have investigated the kinetics of unfolding and folding in a DNA hairpin at different temperatures using the PM and CFM in hopping and pulling experiments. To the best of our knowledge, the implementation of a CFM in pulling experiments with optical tweezers has not yet been reported. It must be said, though, that in other setups such as magnetic and acoustic tweezers the CFM remains as the natural pulling protocol. Using the optical setup described in [7], we carried out hopping experiments using the PM and CFM at three different temperatures over a broad range (6º, 25º and 45ºC) to take advantage of the strong dependence of kinetic rates on temperature. The large change in kinetic rates makes then possible to assess the effect of missed transitions on kinetics measurements in the CFM. This has been done in [18,19] for hopping transitions at room temperature (25º C) and here is extended to low and high temperatures and for pulling experiments too. Experiments in the PM were done at the same three temperatures as for the CFM to cross-check the validity of the CFM results.

## II. MATERIALS AND METHODS

### Preparation of single-molecule construct

The molecular construction has been designed as described in [18]. It was synthesized using the hybridization of two different oligonucleotides. One contains the designed 20 base pair (bp) DNA hairpin (5'-GCGAGCCATAATCTCATCTG GAAA CAGATGAGATTATGGCTCGC-3') flanked with two identical single-stranded DNA (ssDNA) handles. The other one has the complementary sequence of the DNA handles. After the ligation step we have the hairpin between both 29 bp double-stranded DNA (dsDNA) handles. The molecular construction is tagged at one end with a single biotin and a multiple digoxigenin at the other end.

### Instrument design

The instrument used in this study is a miniaturized counterpropagating optical tweezers setup (see details in [6,10]) focused in a microfluidics chamber to hold a microscopic dielectric bead attached to one end of the molecular construction. The other end is fixed in the tip of a micro-pipette by air suction (*FIG 1a*). One bead is coated with streptavidin whereas the other is coated with anti-dig. Connections to beads with the molecular construction are made through biotin-streptavidin and antigen-antibody bonds, respectively. The distance between the center of the optical trap and the immobilized bead is called λ and is the natural control parameter of this device. The instrument can operate in two temperature-controlled modes: hot and cold. In the hot mode a collimated laser of wavelength 1435 nm is placed coaxially to the trapping beams to heat the medium close to the trap [7]. By tuning the laser power of this heating laser, it is possible to increase the temperature from room temperature (25ºC) to a maximum of 50ºC (hot mode). To operate in the cold mode, we introduce the head of the miniaturized tweezers setup inside a turned off icebox containing 20L of ice to cold down the temperature and keep it constant (4-5ºC). By turning on the heating laser inside the icebox, it is possible to increase the temperature from 5º to 30ºC (cold mode). Overall, the combined hot and cold modes gives an operational temperature range between 5ºC to 50ºC.

To determine the real temperature of the sample at each laser power it is necessary to calibrate it using the Stokes's Law test at different trap powers (*FIG 1b*). By moving the bead hold in the optical trap at different velocities and recording the force, we can estimate changes in the viscosity of the medium from the measured slope in the force-velocity curves. Fitting the measured viscosity to the empirical Vogel equation it is possible to determine the real temperature near the optical trap for each laser power.

**Hopping experiments**

In hopping experiments, the control parameter, λ (PM) or force (CFM), is kept constant while folding and unfolding transitions are monitored (*FIG 1c-e*). In the PM the force jumps at each transition, the value of the jump being proportional to the number of released (unfolding) or retracted (folding) base pairs (*FIG. 1c, d*). Also, it is possible to see in *FIG. 1c* how kinetics changes as a function of temperature: the higher the temperature, the faster the hopping kinetics. In the CFM force feedback acts to keep constant the force. When an unfolding or folding event takes place, the linear feedback algorithm calculates the trap displacement necessary to recover the preset force value (*FIG. 1e*). From these experiments we can recover directly the molecular extension at different forces. For both type of hopping experiments, CFM and PM, a partition method based on Gaussian fitting is used to estimate the lifetime of each state. The corresponding kinetic rate is defined as the inverse of the average lifetime. In the case of the CFM experiments the relationship between the force and the kinetic rates is obtained directly from the data. In the PM the force corresponding to the unfolding (folding) transition rates is defined as the average force measured in the folded (unfolded) state.

**Pulling experiments**

In pulling experiments, the control parameter is repeatedly moved back and forth between a minimum value corresponding to a low force where the molecule is folded to a maximal value corresponding to a large force (*FIG. 1d, g, h*). When the molecule unfolds (refolds) at a certain value of the control parameter a jump is observed in the force signal. In pulling experiments in the CFM, when the molecule unfolds (refolds) the feedback transiently increases the pulling speed to recover the same force before the rip (jump). Changes in the trap distance are observed when the molecule changes its configuration (*FIG. 1h*). Although rips in extension in pulling experiments in the CFM are expected, these are not clearly observed along the force-distance curve due to the limited bandwidth of the feedback that is not sufficiently rapid to follow the transitions.

In order to determine the kinetic rates involved in the melting process we measure the first rupture and recovery forces along the unfolding and folding trajectories, respectively. From the histograms of such forces we can estimate the survival probabilities for the folded and unfolded state. Finally, from the survival probabilities it is possible to calculate the unfolding and folding kinetic rates by dividing the probability by the corresponding force histogram and multiplying it with the experimental loading rate (see [26] for details).

**Bell-Evans (BE) model**

Our molecular construct consists of a 20 bp DNA hairpin inserted between two identical short dsDNA handles of 29 bp. The sequence of the stem of the hairpin is designed for the hairpin to fold and unfold as a two-states system. The molecular free energy landscape can then be modeled as two-states system separated by a kinetic barrier. The unfolding/folding reaction rates of this hairpin can be schematically described by a unimolecular reaction pathway,

$$F \underset{k_{F\leftarrow U}}{\overset{k_{F\rightarrow U}}{\rightleftarrows}} U,$$

where F and U indicate the folded and unfolded state and $k_{F\rightarrow U}$, $k_{F\leftarrow U}$ denote the force-dependent unfolding and folding rates, respectively.,

The free energy landscape shows coexistence between both states close to 15 pN and a ssDNA of released extension around 18 nm using the unified oligonucleotide or unzipping thermodynamic dataset and the elastic response of ssDNA at 25ºC [10,20]. The kinetic rates can be studied using the BE model,

$$k_{F\rightarrow U}(f) = k_0 \exp[\beta f x_{F-TS}] \quad (1)$$

$$k_{F\leftarrow U}(f) = k_0 \exp[\beta(\Delta G - f x_{TS-U})] \quad (2)$$

where $k_0$ is the kinetic attempt frequency of the molecule at zero force; $x_{F-TS}$ is the distance from F to the transition state (TS); $x_{TS-U}$ is the distance from TS to U; $\Delta G$ is the energy difference between F and U and $\beta=1/k_BT$ is the inverse of the product of the Boltzmann constant ($k_B$) and the absolute temperature (T).

**Bandwidth limitations of the PM and CFM**

Two kinds of hopping experiments, PM and CFM (see *FIG. 1e, f*), are presented in this work. In the PM the trap position is kept constant and data acquisition is only limited by the sampling rate of the device, 1 kHz. All transitions occurring in a timescale below 1 ms cannot be followed by the instrument. This sampling rate is large enough to measure molecular transitions that occur in the 0.1–1 second scale (e.g.

at 25ºC). In the CFM the force is kept constant using a feedback to keep constant the distance between the center of the optical trap and the center of the pipette bead. The feedback algorithm works by moving the piezoelectric actuators at a rate of 1 kHz. A typical time lag of the piezoelectric actuators is around 1 ms. Moreover, the algorithm has a delay related to the fact that feedback cannot discriminate whether a change in the force signal is due to thermal noise or to a real transition. This algorithm delay is around 5–10 ms. Given these limitations the CFM experiment will miss short time events. In consequence, a kinetic underestimation takes place at low forces for the unfolding process and at high forces for the folding process as previously reported [19]. The effect leads to an overestimation of the free energy difference and the total molecular extension. The latter is the sum of the distances $x_{F\text{-}TS}$ and $x_{TS\text{-}U}$ obtained from *Eqs*. 1 and 2, $x_{F\text{-}TS}+x_{TS\text{-}U}=x_{F\text{-}U}$.

**III. RESULTS**

**Room temperature (25ºC)**

To determine the molecular extension and energy difference it is necessary to fit *Eqs. 1* and *2* to the experimental data (*FIG 2a*). In *FIG. 2a* are shown the average kinetic rates for a minimum of 4 different molecules obtained from CFM and PM hopping experiments at room temperature. Blue full squares correspond to the folding transition rates obtained from CFM while the red full squares are the unfolding rates for the same experiment. The solid lines correspond to the fits at *Eqs. 1* and *2* for the corresponding transition. In addition, blue and red empty circles correspond to the folding and unfolding kinetic rates for the PM experiment. Dashed lines are the corresponding fits to the kinetic rates (red–unfolding and blue–folding). Looking at *FIG. 2a* we can see that the solid lines have a more pronounced slope as compared to the dashed ones, meaning larger distances $x_{F\text{-}TS}$ and $x_{TS\text{-}U}$ and a larger molecular extension $x_{F\text{-}U}$ in the CFM. Finally, we remark that in the studied force range the behavior of kinetics from CFM and PM are well matched with the expected exponential force dependence so characteristic of the BE model. In *Table 1* are summarized the obtained elastic and energetic parameters from both sets of data. It can be seen a difference in the molecular extension of 2.2 ± 1.5 nm between the PM and CFM experiments. This difference agrees with reported values in previous studies with the same molecule [18,19]. A difference of 12 ± 2 $k_BT$ in ΔG agrees with previous published results too [18].

Pulling experiments in the PM and CFM have been also carried out to determine the kinetic rates of the designed DNA hairpin (*FIG 1d, g, h*). Histograms of the first rupture and recovery forces have been measured with a minimum of 4 different molecules with at least 150 cycles per molecule (*FIG. 2b*). PM experiments are done at 70 nm/s and CFM experiments at an equivalent rate of 6 pN/s. In *FIG. 2b* it is possible to see that the first rupture (red dots) and recovery (blue dots) forces match for pulling experiments in CFM (full squares) and PM (empty circles). We note that the mean rupture force is higher than the mean recovery force as expected. The results for the PM experiments are in agreement with reported ones [22]. However, the pulling experiments in the CFM are new results that have not been reported previously, to the best of our knowledge.

From those histograms are recovered the kinetic rates as a function of the force for both experiments (*FIG. 2c*). In *FIG. 2c* are used the same color and symbol criteria than in *FIG. 2b* to denote the CFM and PM experiments and the folding or unfolding kinetic rates. We notice that the kinetic rates for the PM and CFM are compatible. The experimental results are fitted to *Eq. 1* and *Eq. 2* to determine the molecular extension and energy difference. In *Table 1* are summarized the more relevant parameters from those fits. Going in deep, we have recovered: 1) the kinetic rate at the coexistence force, which is close to the one obtained from hopping experiments (0.9 ± 0.2 $s^{-1}$ in pulling and 0.7 ± 0.2 $s^{-1}$ in hopping); 2) the sum of $x_{F\text{-}TS}$ and $x_{TS\text{-}U}$ yields the molecular extension with a difference between CFM and PM of only 0.1 ± 1.4 nm; and 3) the total free energy difference between F and U is 0.6 ± 1.9 $k_BT$ between both modes. Interestingly, the results of pulling experiments in the CFM do not overestimate distances and free energies as observed in CFM hopping experiments.

**B. Cold temperature (6ºC)**

At low temperatures there is much less thermal noise, the unzipping process exhibiting slower kinetics. In addition, lower temperatures contribute to preserve the molecular interaction between the beads and the handles thereby facilitating the experiments. In *FIG. 3* are shown the kinetic results from hopping and pulling experiments with the same symbol and color code than at room temperature. Kinetic rate measurements from hopping experiments carried out in PM and CFM (*FIG. 3a*) are shown to be compatible. The lower the temperature, the lower the strength of the thermal forces and the higher the value of the coexistence force, as expected. The experimental parameters obtained from fitting data to *Eqs. 1* and *2* are summarized in *Table 1*. The values of the change in molecular extension $x_{F\text{-}U}$ and the free energy of formation $\Delta G$ measured in the CFM and PM are compatible within errors.

Pulling experiments at 6 pN/s (CFM) and 70 nm/s (PM) have been also carried out. A minimum of 4 different molecules and 150 trajectories per molecule were collected. As we can see in *FIG. 1d* the hysteresis between the unfolding and folding trajectories increases with respect to room temperature measurements, *FIG. 3b*. As shown in *FIG. 3c*, the unfolding and folding rates obtained using the PM and CFM match perfectly throughout the force range. Let us note that while the coexistence force in pulling and hopping (in both the PM and CFM) are compatible with each other (17.7 ± 0.6 pN) the range of forces where kinetic rates are measured is much larger in the pulling case (see *FIGs. 3a,3c)*. A curvature in the force-dependent kinetic rates is apparent in *FIG*. 3c as predicted by the Leffler-Hammond postulate (26). The parameters fitting Eqs.1 and 2 to the experimental data shown in *FIGs. 3c* are shown in *Table 1*. The values of the change in molecular extension $x_{F\text{-}U}$ and the free energy of formation $\Delta G$ measured in the CFM and PM are compatible within errors. They are also compatible to the values obtained from hopping experiments.

**C. Hot temperature (45ºC)**

Hopping experiments at high temperatures show faster kinetic rates and lower coexistence forces corresponding to larger thermal forces (*FIG. 1c*). We also expect, for CFM data, an increased fraction of missed transition events that should lead to a larger overestimation of both the change in molecular extension $x_{F-U}$ and the free energy of formation $\Delta G$ as compared to the room temperature case. The question remains whether such overestimation is also observed in pulling experiments, either in the PM or in the CFM.

In contrast to the cold case larger forces are required to unzip the DNA hairpin (*FIG. 1d),* a minimum of 5 different molecules with at least 100 pulling trajectories per molecule have been studied. The kinetic measured from hopping experiments are shown in *FIG. 4a*. In this plot the CFM points are presented as full squares, red for the unfolding and blue for the folding, while the points for the PM are denoted as empty circles with the same color criteria. As expected the results obtained from CFM and PM differ due to missed transitions in the CFM. The overestimation of the molecular extension $x_{F-U}$ and free energy difference ΔG are 3.2 ± 1.3 nm and 12 ± 2.5 $k_BT$ respectively. All the parameters obtained from the fitting Eqs.1 and 2 to the kinetic rates are summarized in *Table 2*.

Pulling experiment were also carried out at two different loading and pulling rates in the CFM and PM. In CFM the two loading rates are 6 and 16 pN/s (full and empty squares, respectively, in *FIG. 4b, c)*. In *FIG. 4b* are presented the histograms of the first rupture (red) and recovery (blue) force in the CFM. As expected the difference between the mean rupture and mean recovery force increases when the speed is increased. This agrees with previous PM experiments of the same hairpin at room temperature [22]. From those histograms we have determined the kinetic rates in the CFM (*FIG. 4c)* that are compatible for both loading rates. Interestingly, the curvature in the log(k) vs. force plots is more apparent at this high temperature revealing a larger shift of the position of the transition state with force [26]. The solid and dashed lines are fits of Eqs.1 and 2 to the experimental data at 6pN/s and 16 pN/s respectively. Fits were done around the coexistence force value where a linear relationship between log(k) and force is observed. In *Table 2* are presented some of the obtained kinetic parameters of the fits. The values of the molecular extension $x_{F-U}$ and free energy difference ΔG in the PM hopping experiments and the CFM pulling experiments are in good agreement. Therefore, no overestimation of relevant kinetic parameters from pulling experiments in the CFM is observed.

Pulling experiments in the PM were done at the equivalent pulling speeds of 70 and 225 nm/s. At both speeds an unexpected result was found which is more striking at 70 nm/s. At this speed the histograms along the force axis of the first rupture and recovery force distributions appears exchanged as if the dissipated work was negative (rather than positive), apparently against the second law (full circles in *FIG. 5a*). Looking in detail to *FIG. 1d* we can see 5 different unfolding and folding trajectories where the molecule hops many times before it reaches the final state. The fact there are many transitions along a single pulling curve means that the first rupture force is not indicative of the overall dissipation and that the average dissipated work also gets contributions from multiple unfolding-folding events. In fact, if the pulling speed is increased to 225 nm/s (empty circles) both histograms overlap at the same force regime approaching the standard regime where first rupture forces are typically larger than first folding forces. Interestingly, the folding force distributions do not change with pulling speed whereas the unfolding force distributions do. Kinetic rates are shown in *FIG. 5b* where a curvature for the kinetic rate vs force plots is not seen. The fitting parameters for both pulling speeds are summarized in *Table 2*. The values of the measured molecular extension $x_{F\text{-}U}$ and free energy difference ΔG agree with those reported from PM hopping experiments and CFM pulling experiments. Only hopping experiments in the CFM overestimate such values.

## IV. DISCUSSION AND CONCLUSIONS

Force spectroscopy studies are often done in two standard experimental modes, the constant force mode (CFM) and the passive mode (PM). In the CFM in optical tweezers or AFM the force is controlled using force feedback. In the PM, the position of the force actuator (optical trap in optical tweezers, cantilever in AFM) is controlled, however force fluctuates due to thermal noise or changes due to conformational transitions. Moreover, single molecule experiments are often carried out in equilibrium conditions (hopping experiments) or out-of-equilibrium (pulling experiments). Both operational modes (CFM and PM) and both kind of experiments (hopping and pulling) are suitable to investigate the kinetics of molecular folders at the single molecule level. In hopping experiments, the CFM has been traditionally preferred in order to directly measure force-dependent kinetic rates, whereas in pulling experiments the PM is always used. The main advantage of the CFM in hopping experiments is the automated correction for drift effects in the force. In contrast such effects are always present in the PM resulting in noisier traces and larger errors in kinetic rate measurements. Although the use of dual traps has much alleviated this problem, most optical tweezers setups still use the single trap configuration where force drift in the PM is unavoidable. In contrast, the main disadvantage of the CFM is the finite bandwidth of the feedback that introduces undesired noise effects and delays in the measurements that often lead to missed hopping transitions and overestimation of kinetic rates and free energy differences [19]. In pulling experiments where the trap position (optical tweezers) or cantilever height (AFM) position is continuously changed and force varied between a minimum and a maximum value, force drift effects are not so acute as in hopping experiments, and the need of the CFM appears less justified. Therefore, pulling experiments in optical tweezers and AFM are always carried out in the PM.

In this paper we have investigated the performance of the CFM and PM in both hopping and pulling experiments. Previous studies have shown limitations of the CFM in hopping experiments related to missing fast transitions due to the finite response time of the feedback. As shown in [19] this effect should therefore be more pronounced in systems with faster kinetic rates. However, it remains unclear whether such limitation is encountered in nonequilibrium pulling experiments in the CFM where irreversibility and dissipation effects effectively decrease the overall fraction of missed transitions.

In this paper we have carried out force spectroscopy experiments on a DNA hairpin at three different temperatures: room temperature (25ºC), a cold temperature (6ºC) and a hot temperature (45ºC) measuring the unfolding and folding kinetic rates in hopping and pulling experiments in both CFM and PM. In general, we find that pulling experiments in both the CFM and PM recover the correct values of the kinetic rates and the corresponding parameters of the free energy landscape, such as molecular extension and free energy of formation. In contrast, hopping experiments yield the correct parameters only in the PM whereas in the CFM the values of the molecular extension of the hairpin and the free energy are overestimated due to missed transitions [19]. This effect is noticeable at room temperature, negligible at low temperature (where hopping kinetics is slower) and dramatic at large temperature (where hopping is faster). Our results, are summarized in *Tables 1* and *2*. More specifically: 1) the molecular extension at room temperature (25ºC) from CFM is 12% larger than the expected one from PM, which agrees with the results in [18,19]. 2) At low temperature (6ºC) this difference is on the same order of magnitude of our experimental error. 3) The free energy of formation ΔG measured at 25ºC using the CFM is 18% higher than the measured one with PM. 4) The values of ΔG measured at 6ºC using CFM an PM are compatible within the experimental errors. In addition, we carried out hopping experiments at 45ºC and confirmed that for high kinetic rates, higher than at room temperature, the feedback is inefficient with many missed transitions and large overestimation effects in the CFM (*Table 2)*: 1) the molecular extension measured from CFM is 20% larger than the one measured from PM experiments. 2) The ΔG measured in CFM is 25% higher than in PM.

Moreover, pulling experiments in a DNA hairpin controlling the trap position (PM) and the force (CFM) have been done at the same temperatures. This is the first time that both kind of pulling experiments, CFM and PM, are done and compared in this temperature range. We have shown that pulling experiments in the CFM and PM yield the correct force-dependent kinetic rates at the three temperatures. The results are summarized in *Tables 1* and *2*. From those results we confirm that both kind of pulling modes, CFM and PM, gives us compatible molecular extensions and free energy differences at each temperature. In addition, pulling experiments present advantages compared to hopping experiments: 1) A larger force range to extract the kinetic rates is covered in pulling experiments. 2) Pulling experiments are less influenced by force drift by aligning individual trajectories in the data analysis. 3) Irreversible pulling experiments result in a faster experimental protocol than equilibrium hopping experiments. 4) Data obtained from irreversible pulling experiments allow us to use other methods, *e.g.* the fluctuation theorem, to extract the free energy of formation of the hairpin. Moreover, we have verified the presence of a concave curvature in the log(k) versus force plots as predicted by the Leffler-Hammond postulate [26]. This concavity is more visible in CFM than in PM data.

Interestingly a new effect has been observed in pulling experiments in the PM at 45ºC and slow pulling rates, 70 nm/s, where the mean first unfolding force is lower than mean first recovery force, the unfolding force distribution being shifted to the left of the recovery force distribution. This result may look as counter-intuitive, and seemingly against the second law. However, it is not as the average dissipated work along the pulling cycle is still positive if one considers all hopping events during the pulling cycle. In *FIG. 1d* one observes the high number of hopping events at 45ºC along single pulling trajectories. The work done upon the hairpin (e.g. along the stretching part of the cycle) equals the area below the force-distance curve, yet this work gets contributions from the multiple transition events along the curve rather than just the first force jump. To confirm that this effect is only related to the low pulling speed we have done experiments at 225 nm/s. At this velocity we have observed that both histograms approach each other overlapping over the same range of forces. The ¨normal ordering¨ of force distributions is then recovered at larger pulling speeds. It is interesting to note that this reversal effect of the first unfolding and folding force distributions allows us to extract kinetic rates over a very broad range of forces, in particular it allows us to determine kinetic rates below the coexistence rate value.

Finally, we have found that in irreversible pulling experiments in the CFM the effect of overestimating distances and energies is less critical than in hopping experiments. Although the feedback algorithm in pulling experiments suffers from the same finite bandwidth limitations as in hopping experiments, the fraction of missed transitions is decreased for the pulling case under irreversible conditions where the first unfolding and recovery transition event is stabilized by the fast increasing (unfolding) or decreasing (folding) force. This is only true in cases where the average first unfolding (folding) force is larger (smaller) than the coexistence force, i.e. under nonequilibrium or irreversible conditions where average dissipation is larger than $k_BT$. Otherwise, if dissipation is low the pulling protocol becomes equivalent to the hopping protocol rendering the CFM inefficient too.

Overall we have confirmed that the overestimation error committed in the determination of the molecular extension and ΔG using the CFM in hopping experiments comes from missed transitions (19), and that this effect tends to increase with temperature as expected. In contrast irreversible pulling experiments in the CFM lead to more reliable results. We conclude that irreversible pulling experiments are a valuable tool to determine the force-dependent kinetic rates in the temperature range 5º-50ºC.

## ACKNOWLEDGMENTS

All authors acknowledge Spanish Research Council Grant FIS2016-80458-P, European Union's Horizon 2020 grant No 687089 and ICREA Academia 2013.

**TABLES**

Table 1. Results from the linear fits of the log(k) versus force for CFM and PM experiments at room and cold temperatures.

| | $f_c$ (pN) | $k_c$ ($s^{-1}$) | ΔG ($k_BT$) | $x_{F-TS}$ (nm) | $x_{TS-U}$ (nm) | $x_{F-U}$ (nm) |
|---|---|---|---|---|---|---|
| Room temperature | | | | | | |
| Hop. CFM | 15.2 ± 0.5 | 0.7 ± 0.2 | 77 ± 1.5 | 10.3 ± 0.5 | 10.1 ± 0.6 | 20.5 ± 1.1 |
| Hop. PM | 15.2 ± 0.3 | 0.7 ± 0.2 | 65 ± 1.2 | 10.3 ± 0.5 | 8.0 ± 0.5 | 18.3 ± 0.9 |
| Pul. CFM | 15.2 ± 0.3 | 0.9 ± 0.2 | 63 ± 1.3 | 8.5 ± 0.5 | 8.7 ± 0.6 | 17.2 ± 1.1 |
| Pul. PM | 15.0 ± 0.3 | 1.0 ± 0.2 | 63 ± 1.4 | 8.6 ± 0.5 | 8.7 ± 0.4 | 17.3 ± 0.6 |
| Cold temperature | | | | | | |
| Hop. CFM | 17.7 ± 0.3 | 0.2 ± 0.1 | 84 ± 1.5 | 8.6 ± 0.8 | 9.8 ± 0.6 | 18.2 ± 1 |
| Hop. PM | 17.7 ± 0.2 | 0.2 ± 0.1 | 82 ± 1.1 | 10.2 ± 0.6 | 7.6 ± 0.6 | 17.8 ± 0.8 |
| Pul. CFM | 17.6 ± 0.3 | 0.3 ± 0.1 | 78 ± 1.8 | 9.1 ± 0.6 | 8.0 ± 0.5 | 17.1 ± 1.1 |
| Pul. PM | 17.5 ± 0.3 | 0.3 ± 0.1 | 76 ± 1.6 | 9.0 ± 0.5 | 7.7 ± 0.6 | 16.7 ± 1.1 |

‡ The presented values are the mean ± standard deviation over 4 different molecules.

† Hop. is hopping and Pul. is pulling

Table 2. Results from the linear fits of the log(k) versus force for CFM and PM at hot temperature.

| | Hot temperature | | | | | |
|---|---|---|---|---|---|---|
| | $f_c$ (pN) | $k_c$ ($s^{-1}$) | ΔG ($k_BT$) | $x_{F-TS}$ (nm) | $x_{TS-U}$ (nm) | $x_{F-U}$ (nm) |
| Hop. CFM | 12.6 ± 0.3 | 6 ± 2 | 59 ± 1.5 | 9.5 ± 0.5 | 10 ± 1 | 19.6 ± 1.1 |
| Hop. PM | 12.6 ± 0.2 | 5 ± 2 | 47 ± 1.5 | 8.1 ± 0.5 | 8.3 ± 0.5 | 16.4 ± 0.7 |
| Pulling Experiments | | | | | | |
| 6 pN/s | 12.4 ± 0.3 | 4 ± 2 | 52 ± 2 | 8.0 ± 0.5 | 8.3 ± 0.6 | 16.3 ± 1.1 |
| 16 pN/s | 12.4 ± 0.3 | 4 ± 1.9 | 46 ± 2 | 8.7 ± 0.8 | 7.7 ± 0.6 | 16.4 ± 1.1 |
| 70 nm/s | 12.5 ± 0.2 | 31 ± 10 | 47 ± 1.8 | 8.1 ± 0.5 | 9.0 ± 0.4 | 17.0 ± 0.7 |
| 225 nm/s | 12.9 ± 0.2 | 16 ± 7 | 49 ± 2 | 8.0 ± 0.5 | 8.6 ± 0.5 | 16.6 ± 0.7 |

‡ The presented values are the mean ± standard deviation over 4 different molecules.

† Hop. is hopping and Pul. is pulling

**FIGURES**

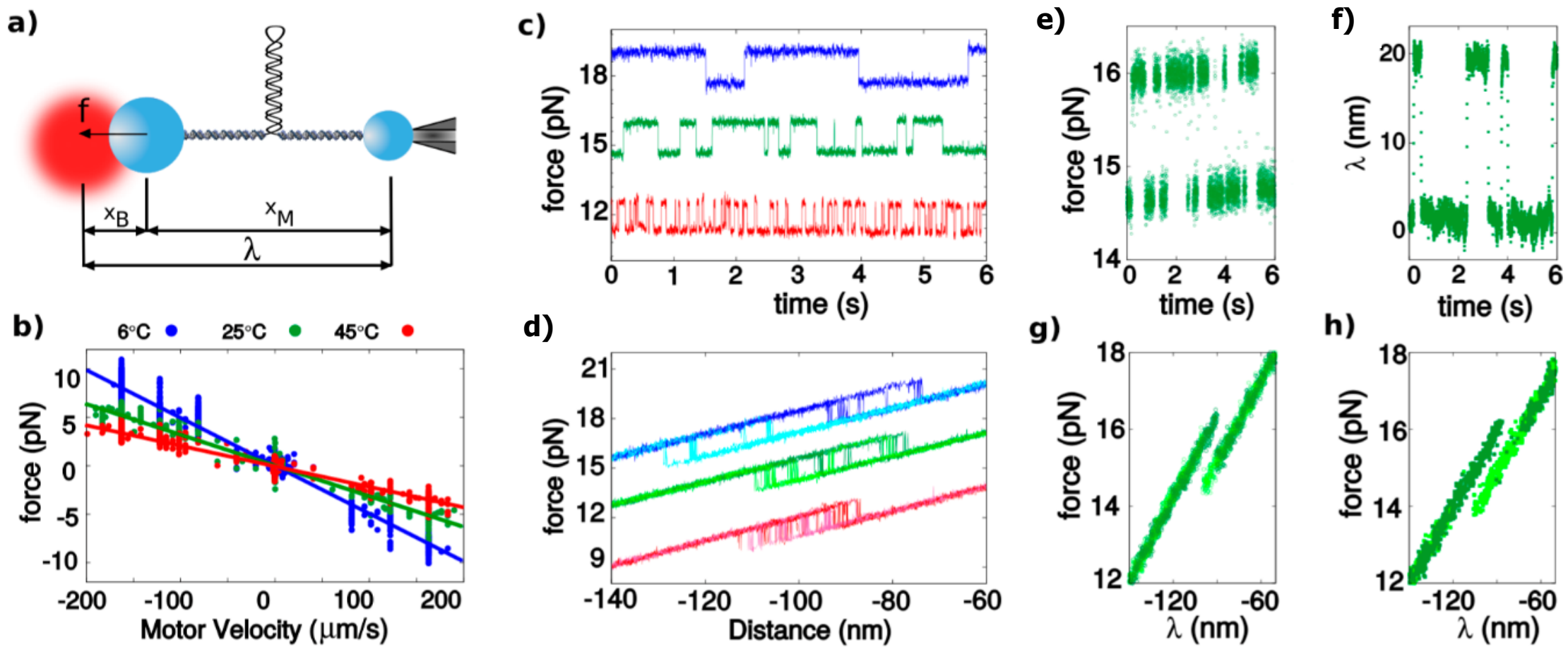

FIG. 1. a) Schematic depiction of the experimental optical tweezers setup to manipulate a single molecule. b) Force versus velocity curve at three different heating powers. Blue points correspond to 6ºC, green points to 25ºC and red ones to 45ºC. Hopping (c) and pulling (d) experiments in PM same color code as in b). Hopping experiments in PM (e) and CFM (f) at 25ºC. f) Pulling experiments in PM (g) and CFM (h) at 25ºC.

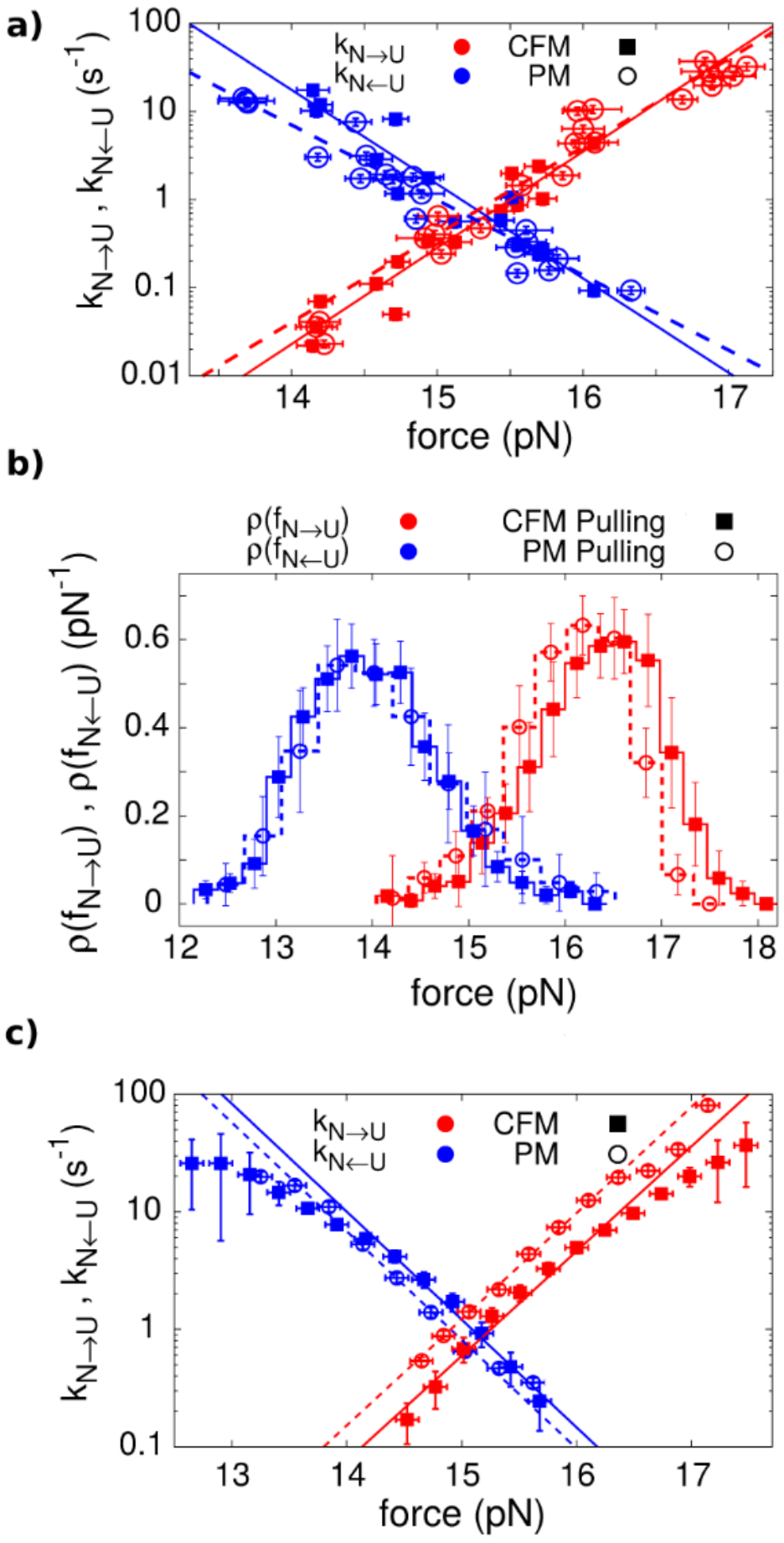

FIG. 2. a) Unfolding kinetic rates (red dots) and folding kinetic rates (blue dots) for the PM (empty circles) and CFM (full squares) for hopping experiments at 25ºC. b) Histograms of first rupture (red dots) and recovery force (blue dots) for PM (empty circles) and CFM (full squares) from pulling experiments at 25ºC. c) Unfolding kinetic rates (red dots) and folding rates (blue dots) for PM (empty circles) and CFM (full squares) from pulling experiments at 25ºC.

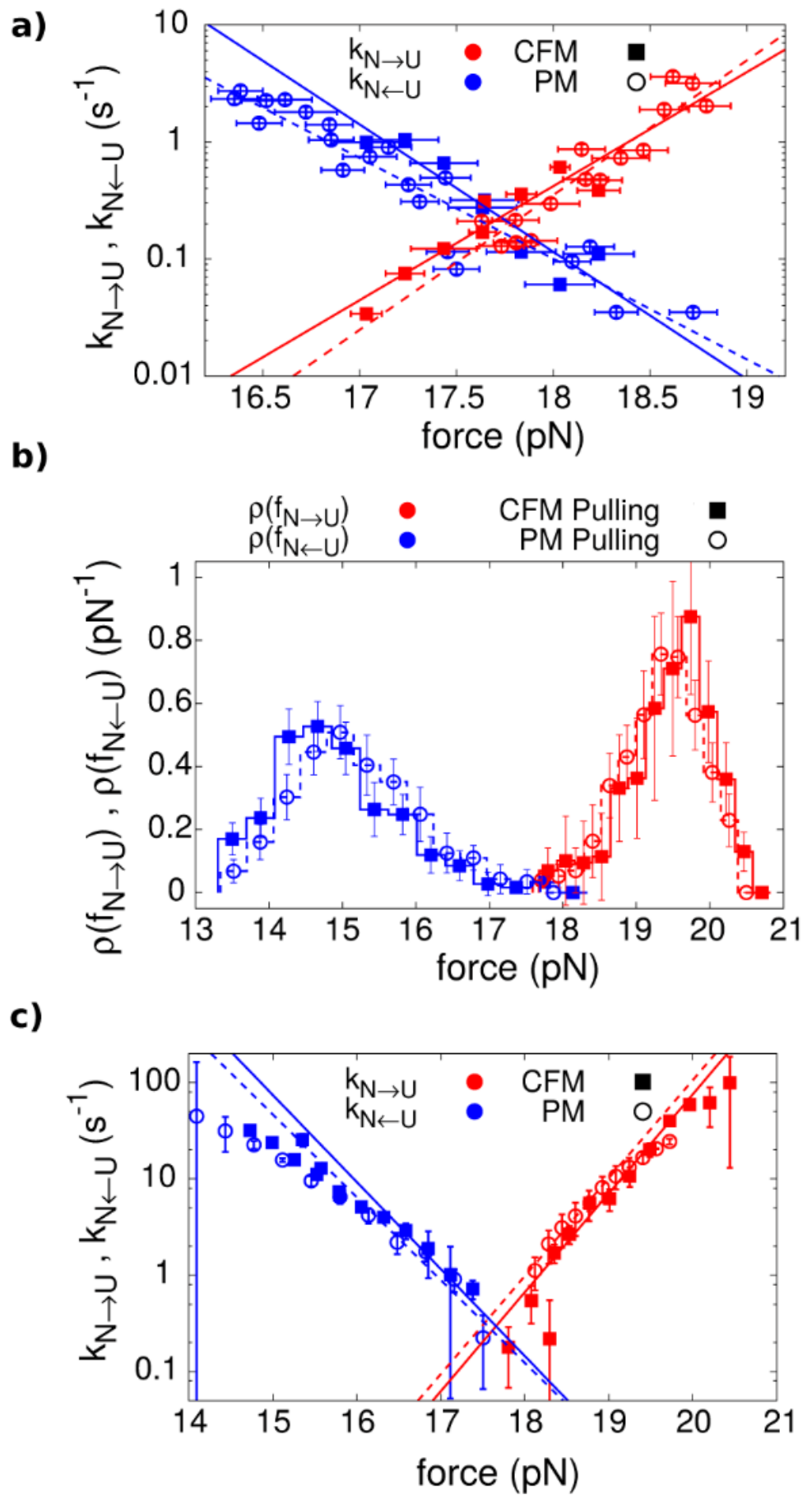

FIG. 3. a) Unfolding kinetic rates (red dots) and folding kinetic rates (blue dots) for PM (empty circles) and CFM (full squares) from hopping experiments at 6ºC. b) Histograms of first rupture (red dots) and recovery force (blue dots) for PM (empty circles) and CFM (full squares) from pulling experiments at 6ºC. c) Unfolding kinetic rates (red dots) and folding rates (blue dots) for PM (empty circles) and CFM (full squares) from pulling experiments at 6ºC.

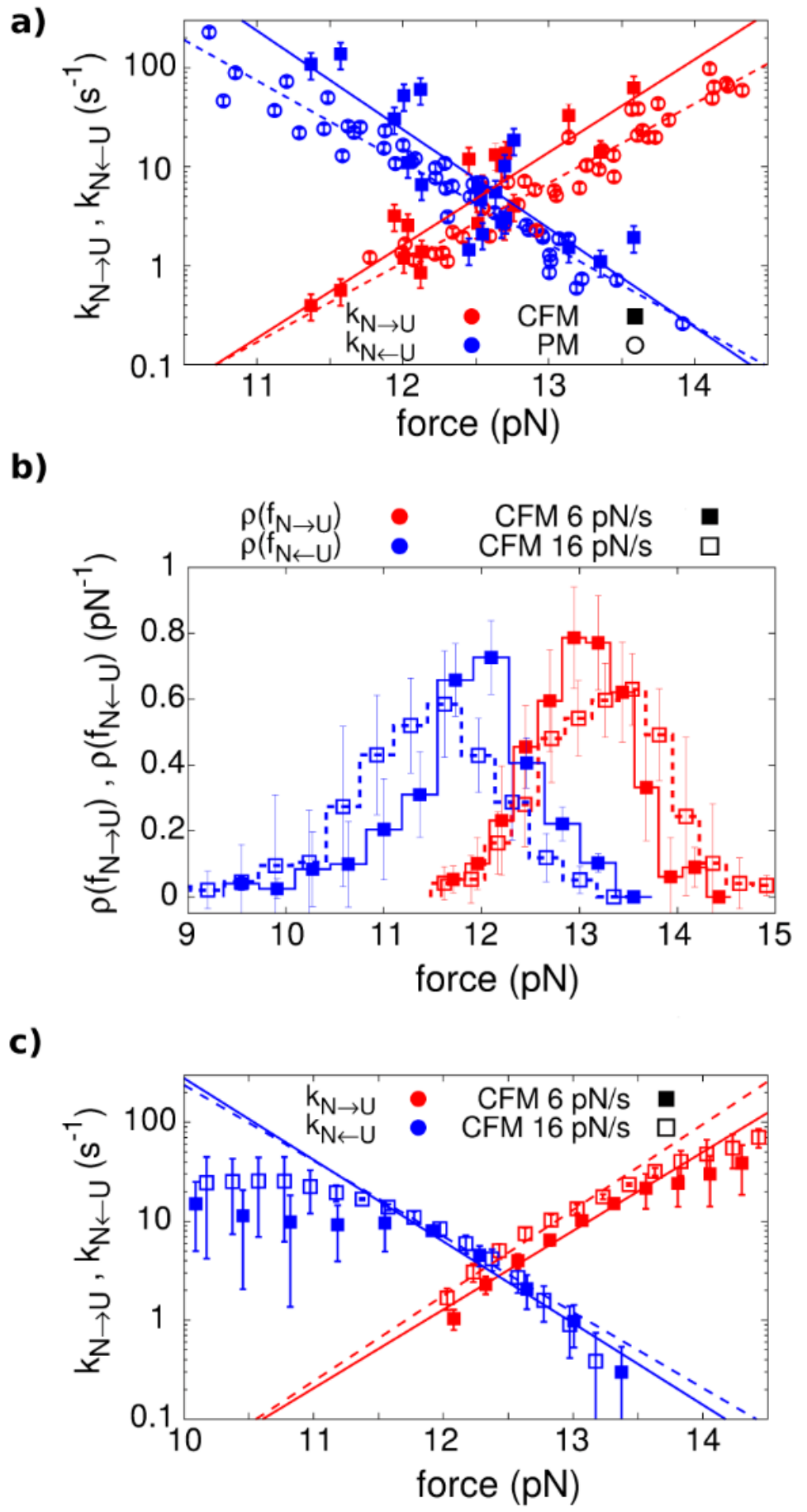

FIG. 4. a) Unfolding kinetic rates (red dots) and folding kinetic rates (blue dots) for PM (empty circles) and CFM (full squares) from hopping experiments at 45ºC. b) Histograms of first rupture (red dots) and recovery force (blue dots) for the CFM pulling experiments at 6 pN/s (full squares) and 16 pN/s (empty squares) at 45ºC. c) Unfolding kinetic rates (red dots) and folding rates (blue dots) for the CFM pulling experiments at two different loading rates, 6 pN/s (full squares) and 16 pN/s (empty squares) at 45ºC.

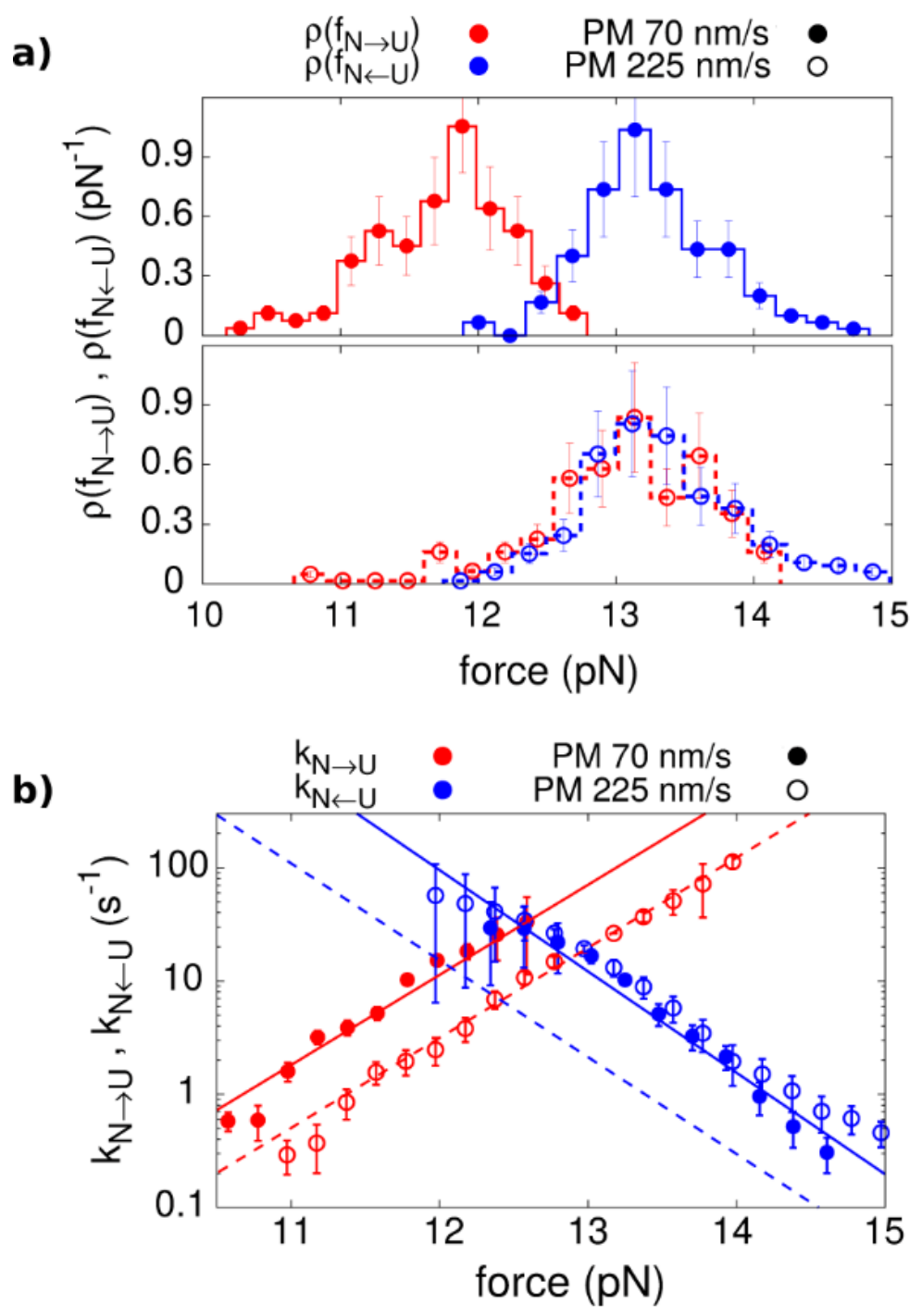

FIG. 5. a) Histograms of first rupture (red dots) and recovery force (blue dots) for the pulling PM experiments at two pulling rates, 70 nm/s (full circles, Top) and 225 nm/s (empty circles, Bottom), at 45ºC. b) Unfolding kinetic rates (red dots) and folding rates (blue dots) at 45ºC for PM pulling experiments at 70 nm/s (full circles) and 225 nm/s (empty circles).